\documentstyle[12pt,epsfig]{article}
\begin {document}

\begin{center}
{\bf RATIOS OF ANTIBARYON/BARYON YIELDS IN HEAVY ION COLLISIONS}

\vspace {2cm}

Yu.M.Shabelski \footnote{E-mail SHABELSK@THD.PNPI.SPB.RU} \\
Petersburg Nuclear Physics Institute, \\
Gatchina, St. Petersburg 188300 Russia \\
\end{center}
{\it Talk, given at Conf. "New Worlds in Astroparticle Physics", \\
5-7 September 2002, Faro, Portugal}  \\

\vspace {1.5cm}

\begin{abstract}
We discuss the model predictions for antibaryon/baryon production ratios in
high energy heavy ion collisions. The role of string junction mechanism 
of baryon number transfer seems to be very important here and we consider 
some quantitative results.
\end{abstract}

\vspace {1.cm}

The first RHIC experimental data collected in \cite{1} show that the 
values of antibaryon/baryon production ratios in midrapidity region 
of Au-Au central collisions at $\sqrt{s_{NN}} = 130$ GeV are rather 
small in comparison with the most of theoretical predictions. Really, the
values of the order of 0.6 for $\bar{p}/p$ and $0.73 \pm 0.03$ for
$\bar{\Lambda}/\Lambda$ were measured whereas the standard Quark-Gluon
String Model (QGSM) \cite{2,3,4,5} predicts in both cases the values more 
than 0.9 and in the String Fusion Model \cite{1} the predicted values are
about 0.8 and about 0.87, respectively.

The main part of $B$ and $\bar{B}$ should be produced at high energies as
$B-\bar{B}$ pairs. It means that the additional source of the baryons in the
midrapidity region is needed. The realistic source of the transfer of baryons
charge over long rapidity distances can be realized via the string junction
(SJ) diffusion where it can combain three sea quarks into a secondary baryon
(but not antibaryon), see \cite{6}.

In such a picture where the transfere of baryon charge is connected with SJ
exchange there exist three different possibilities to produce a secondary
baryon which are shown in Fig. 1 \cite{7}.

This additional transfer of baryon number to the midrapidity region is
governs by contribution \cite{7}
\begin{equation}
G^p_{uu} = G^p_{ud} = a_N \sqrt{z}[v_0 \epsilon (1 - z)^{1 - \alpha_{SJ}}
+ v_q z^{3/2} (1-z) + v_{qq} z^2] \;,
\end{equation}
where the items proportional to $v_{qq}$, $v_q$ and $v_0$ correspond
to the contributions of diagrams Fig. 1a, 1b and 1c, respectively.
The most important in the midrapidity region at high energies is the
diagram Fig. 1c which obeys the baryon number transfer to rather
large rapidity region.

In agreement with the experimental data the parameter $\epsilon$
in Eq. (1) is rather small. However different data are in some
disagreement with each other, see the detailed analyses in
\cite{7}. Say, $x_F$ -distributions of secondary protons and
antiprotons produced at 100 and 175 GeV/c and at $\sqrt{s}$ = 17.3 GeV 
are in better agreement with $\epsilon = 0.05$, whereas ISR data and
the $p/\bar{p}$ asymmetry at HERA energy are described better with 
$\epsilon = 0.2$.

Some part of this disagreement can be connected with different
energies. In Fig. 1c, as a minimum, two additional mesons $M$ should 
be produced in one of the strings, as it is shown in this Fig., 
that can give the additional smallness \cite{11} at not very high 
energy due to decrease the available phase space. Another source of 
disagreement of the data at low and high energies can come from the fact 
that the suppression of baryon number transfer to large rapidity 
distance $\delta y$ should be proportional to
\begin{equation}
e^{-(\alpha_{SJ}-1)\cdot  \Delta y} \:,
\end{equation}
and the effective value of $\alpha_{SJ}$ depends on the energy due to 
Regge cut contribution \cite{12}.
  
In the case of $\pi^-p \to \Omega X, \bar{\Omega} X$ reactions the 
contribution (1) leads to the contradiction with the simplest version of 
additive quark model \cite{13} because experimentally (see \cite{7}) the 
yields of $\Omega$ in the central region is permanently larger than the 
yields of $\bar{\Omega}$.  

Let us note that it is rather dangerous to use large value of $\epsilon$
as well as the value of $\alpha_{SJ}$ close to unity in Eq. (1). The 
reason is that the string junction mechanism can not transfer more 
baryon charge than we have in an incident state. The hadron content of 
sea-quark baryons can be written \cite{CS} as
\begin{equation}
(u + d + \lambda s)^3 = 4p + 4n + 12\lambda (\Lambda + \Sigma) + ... \;,
\end{equation} 
where $\lambda$ is the suppression factor for strange quark production.
So the integral multiplicity of the protons produced via SJ mechanism
from one incident baryon can not be larger than 
$W_p = 4/((8 + 12\lambda) \approx 0.3-0.4$ (for $\lambda = 0.2 - 0.4$) :
\begin{equation}
\epsilon \int_0^{\infty} e^{(\alpha_{SJ} - 1) \Delta y} d \Delta y 
\leq W_p
\end{equation}

The HERA data on $p/\bar{p}$ asymmetry can be described \cite{7} with the 
parameters which are near to the presented boundary, namely 
$\alpha_{SJ} = 0.5 , \epsilon = 0.2$. However, if we use
the large part of the initial baryon charge for its diffusion to the 
mid-rapidity region, the multiplicity of secondary baryons in the 
fragmentation region should significantly decrease. It will result in 
additional mechanism of Feynman scaling violation in the fragmentation 
region in comparison with the estimations which were claimed in 
\cite{14}. 

In the case of hadron-nucleus collisions the yields of secondaries can be
calculated in QGSM by similar way \cite{3,14} and also with including SJ 
contributions.

\begin{figure}[htb]
\centerline{
\mbox{\epsfig{file=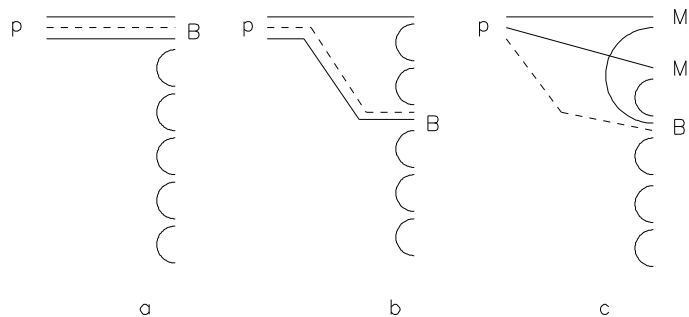,width=0.7\textwidth}}}
\caption{
 Three different possibilities of secondary baryon production in $pp$
collisions: SJ together with two valence and one sea quarks (a), together
with one valence and two sea quarks (b), together with three sea quarks (c).
}
\end{figure}

In the case of heavy ion (A-B) collisions the multiple scattering theory
allows one to account for the contribution af all Glauber-type diagrams
only using some Monte Carlo method where the integrals with dimention of
about $2 \cdot A \cdot B$ should be calculated in coordinate space 
\cite{15,16}. The analitical calculations allow one to account only some 
classes of diagrams \cite{17,18,19}. One of approaches here is so called 
rigid target approximation where it is assumed that for the forward hemisphere we can neglect by the
binding of projectile nucleons (i.e. consider them as a beam of free nucleons
and every of them can interact with target nucleus). The last one is considered
as a dense medium. And vice versa, for the backward hemisphere we consider
the target nucleons as a beam of free nucleons and every of them can interact
with dense medium of projectile nucleus. All details and needed formulae can
be found in \cite{5}. The resulting expression for secondary $h$
production in $A-B$ collisions reads as
\begin{eqnarray}
\frac1{\sigma^{prod}_{AB}} \frac{AB \to hX}{dy} & = &
\theta(y)\langle N_A\rangle \frac1{\sigma^{prod}_{NB}} \frac{NB \to
hX}{dy} +  \nonumber\\
& + & \theta(-y)\langle N_B\rangle \frac1{\sigma^{prod}_{NA}} \frac{NA
\to hX}{dy}\ ,
\end{eqnarray}
where $\langle N_A\rangle$ and $\langle N_B\rangle$ are the average 
numbers of interacting nucleons in nuclei $A$ and $B$.  They depend on 
the $A-B$ impact parameter, $A/B$ ratio, etc. \cite{20}.

The calculated ratios of $\bar{p}/p$ and $\bar{\Lambda}/\Lambda$
production in $Au-Au$ collisions at RHIC, predicted by Eq.(5) with 
accounting for the percolation effects \cite{JShU} are presented in 
Table 1 \cite{Sh}. One can see that small, $\epsilon = 0.05$, SJ 
contribution can not explain the data. Comparatively large contribution
($\epsilon = 0.2$), close to the upper limit (4) gives the ratios more 
close to their experimental values. The more accurate accounting of 
inelastic shadowing/percolations (multipomeron interactions) 
\cite {JShU,21} can lead to better agreement with the data.

\begin{table}
\caption{Antibaryon/baryon yields for secondaries produced at RHIC in
midrapidity region at $\sqrt = 130$ GeV}.
\begin{center}
\begin{tabular}{||c|c|c|c||} \hline \hline
& \multicolumn{2}{c|}{QGSM} &  Exper.  \\  \hline

& $\epsilon = 0.05$ & $\epsilon=0.2$ & \\ \hline

$\bar p/p$ & 0.83  & 0.67  &  $\sim 0.6$  \\ \hline

$\bar\Lambda/\Lambda$ & 0.83 & 0.64 & $0.73\pm0.03$ \\
\hline \hline
\end{tabular}
\end{center}
\end{table}

In conclusion we note that $\bar{B}/B$ asymmetry in midrapidity region
at RHIC energies is rather large. It can be explained by large SJ 
contribution that is in reasonable agreement with HERA data.

\subsection*{Acknowledgements}
This work was supported by grants NATO PSTCLG 977275 and 
RFBR 98-02-17629. I am very grateful to G.Arakelyan, A.Capella,
J.Dias de Deus, A.Kaidalov, C.Pajares and R.Ugoccioni for collaboration 
and useful discussions.

\end{document}